\documentclass[abbrv,aps,prl,twocolumn,showpacs,preprintnumbers,amsmath,amssymb,
superscriptaddress,nofootinbib]{revtex4}

 \pdfoutput=1
\usepackage[T1]{fontenc}
\usepackage[latin1]{inputenc}
\usepackage{lmodern}
\usepackage{graphicx} 
\usepackage{tabularx} 
\usepackage{dcolumn}
\usepackage{bm}
\usepackage{psfrag}
\usepackage{epsfig}
\usepackage{color}
\usepackage{ulem}
\usepackage{amsmath}
\usepackage{amssymb}

\def\ii{{\rm i}}

\makeatletter
\newcommand*{\balancecolsandclearpage}{
  \close@column@grid
  \clearpage
  \twocolumngrid
}
\makeatother

\begin{document}
\title {Plasmon-phonon interactions in topological insulator rings}

\author{M.~Autore} 
\affiliation{INFN and Dipartimento di Fisica, Universit\`a di Roma "La Sapienza",
Piazzale A. Moro 2, I-00185 Roma, Italy}
\author{F.~D'Apuzzo}
\affiliation{Istituto Italiano di Tecnologia and Dipartimento di Fisica, Universit\`a di Roma La Sapienza, Piazzale Aldo Moro 2, I-00185 Roma, Italy}
\author{A.~Di Gaspare} 
\affiliation{CNR-IFN and LNF-INFN, Via E. Fermi 40, 00044 Frascati, Italy}
\author{V.~Giliberti} 
\affiliation{CNR-IFN and Dipartimento di Fisica, Universit\`a di Roma "La Sapienza",
Piazzale A. Moro 2, I-00185 Roma, Italy}
\author{O.~Limaj} 
\affiliation{INFN and Dipartimento di Fisica, Universit\`a di Roma "La Sapienza",
Piazzale A. Moro 2, I-00185 Roma, Italy}
\author{P.~Roy} 
\affiliation{Synchrotron Soleil, L'Orme des Merisiers, Saint Aubin, BP 48, 91192 Gif-sur-Yvette, France}
\author{M.~Brahlek} 
\affiliation{Department of Physics and Astronomy Rutgers, The State University of New Jersey 136 Frelinghuysen Road Piscataway, NJ 08854-8019 USA}
\author{N.~Koirala} 
\affiliation{Department of Physics and Astronomy Rutgers, The State University of New Jersey 136 Frelinghuysen Road Piscataway, NJ 08854-8019 USA}
\author{S.~Oh}
\affiliation{Department of Physics and Astronomy Rutgers, The State University of New Jersey 136 Frelinghuysen Road Piscataway, NJ 08854-8019 USA} 
\author{F.~J.~Garc\'{\i}a~de~Abajo*}
\affiliation{ICFO-Institut de Ciencies Fotoniques, Mediterranean Technology Park, 08860 Castelldefels (Barcelona), Spain}
\altaffiliation{ICREA-Instituci\'o Catalana de Recerca i Estudis Avan\c{c}ats, Passeig Llu\'{\i}s Companys 23, 08010 Barcelona, Spain}
\author{S.~Lupi*}
\affiliation{INFN, CNR-IOM, and Dipartimento di Fisica, Universit\`a di Roma "La Sapienza",
Piazzale A. Moro 2, I-00185 Roma, Italy}
\date{\today}


\maketitle

\textbf{The great potential of Dirac electrons for plasmonics and photonics has been readily recognized after their discovery in graphene, followed by applications to smart optical devices. Dirac carriers are also found in topological insulators (TI) --quantum systems having an insulating gap in the bulk and intrinsic Dirac metallic states at the surface--. Here, we investigate the plasmonic response of ring structures patterned in Bi$_2$Se$_3$ TI films, which we investigate through terahertz (THz) spectroscopy. The rings are observed to exhibit a bonding and an antibonding plasmon modes, which we tune in frequency by varying their diameter. We develop an analytical theory based on the THz conductivity of unpatterned films, which accurately describes the strong plasmon-phonon hybridization and Fano interference experimentally observed as the bonding plasmon is swiped across the promineng 2\,THz phonon exhibited by this material. This work opens the road for the investigation of plasmons in topological insulators and for their application in tunable THz devices.}

Plasmons --the collective oscillations of charge carriers in conducting materials-- hold great potential for combining electronics and photonics at the nanoscale. These excitations can propagate along extended surfaces or they can be localized at so-called hotspots. The engineering of plasmon frequencies and spatial profiles has been mastered in metallic micro- and nanostructures such as nanospheres, nanorods \cite{LE99}, dimers \cite{NOP04,GRP05}, particle arrays \cite{L06,TPP98}, and a plethora of more exotic morphologies \cite{ASH97,paper042,paper097,GPM08}. Plasmon hybridization \cite{PRH03,FTS13}, Fano resonances \cite{LZM10,MFK10}, and electromagnetically induced transparency \cite{YBCO-Odeta} are among the feats that have been realized and broadly used to understand and design plasmonic devices. The range of applications of plasmon excitations is vast and includes optical sensing \cite{KE06,AHL08, Plasmonics1, APL}, quantum electrodynamics \cite{CSH06,AMY07}, nonlinar optics \cite{DN07,DSK08}, photovoltaic technologies \cite{AP10}, and medical diagnosis and treatment \cite{NHH04,JEE07}.

Extensive experimental efforts are currently underway to find materials with improved plasmonic performance, in particular in the mid-infrared and terahertz parts of the electromagnetic spectrum. Examples of such materials are low- \cite{LT} and high-Tc superconductors \cite{YBCO-Odeta}, conductive oxides \cite{ITO}, and graphene \cite{WSS06,BOS07,FAB11,JGH11,FRA12,paper212,BJS13}. The latter exhibits a peculiar electronic structure, which enables unprecedented levels of electro-optical tunability via chemical or electrostatic doping \cite{NGM04,NGM05,MSW08}: electrons in graphene behave as massless Dirac fermions characterized by a linear dispersion relation and a vanishing density of states at the Fermi energy, so that a few additional charge carriers produce a large shift in the chemical potential. Dirac charge carriers are also found in three-dimensional topological insulator materials --i.e., quantum systems characterized by an insulating electronic gap in the bulk, which opening is due to strong spin-orbit interaction, and gapless surface states at the edge and interface. Surface states in TIs are metallic and chiral, associated with massless Dirac quasi-particles \cite{Hasan,KaneMele,Moore}, and protected from back-scattering by time-reversal symmetry. This means that they cannot be destroyed or gapped by scattering processes, other than magnetic impurities. Furthermore, unlike graphene, TI surface states spontaneously provide a 2D Dirac-fermion system, segregated from the bulk material without the need of physically implementing an atomic monolayer. Since their discovery,  TI's have attracted a growing interest due to their potential application in quantum computing \cite{Fu1,Kitaev1}, THz detectors \cite{zangh}, and spintronic devices \cite{Chen}, and some of these applications could benefit from propagating and localized plasmons sustainted by these materials. 

Recently, propagating plasmons have been observed in the THz domain in micro-ribbon arrays patterned on Bi$_2$Se$_3$ TI films \cite{NNANO}. The plasmon frequency $\nu_P$ was found to follow an optical dispersion relation $\nu_P\propto1/\sqrt{W}$, where $W$ is the ribbon width, characteristic of 2D metal systems and in excellent agreement with that predicted for Dirac plasmons. \cite{Polini,DasSarma} However, despite their potential for novel applications, neither localized plasmons nor their hybridization and coupling have yet been explored in TI micro and nano structures. In this Letter, we reveal the behavior of localized plasmons in topological insulator rings (TIRs), with a particular focus on their dependence on size and temperature, as well as on their hybridization with a Bi$_2$Se$_3$ optical phonon located at 2 THz. We demonstrate geometrical tuning of the TIR plasmons and the resulting hybridization that exhibit a strong temperature dependence as observed in transmission spectra. This study indicates that TIRs hold a great potential for applications in thermally controlled THz devices.

\begin{figure*}
\begin{center}
\includegraphics[width=130mm,angle=0,clip]{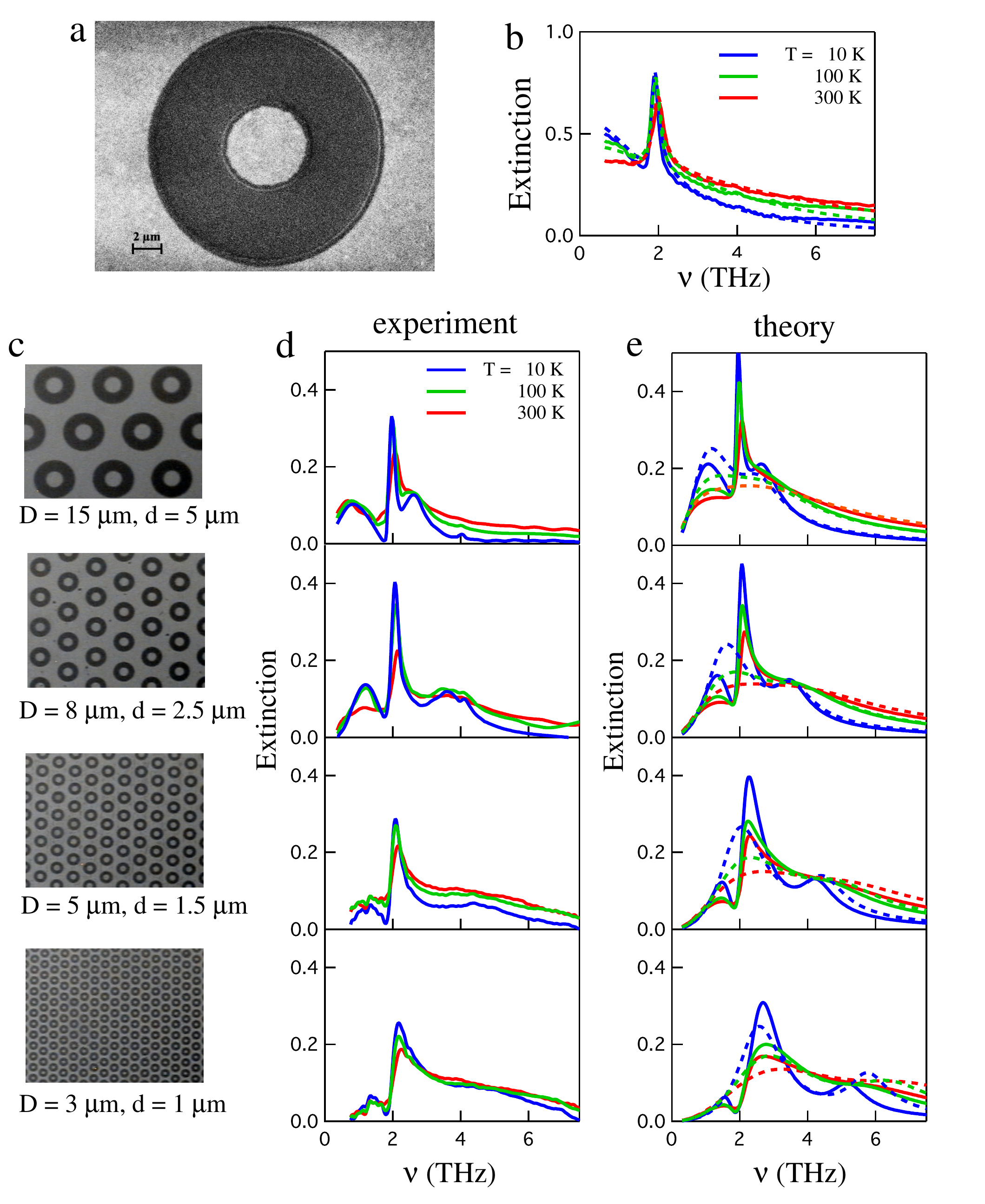}
\caption{{\bf Geometrical tuning of the plasmonic response of topological insulator rings (TIRs).} {\bf (a)} Detailed SEM image of a Bi$_2$Se$_3$ TIR having an outer (inner) diameter $D$= 15 ($d=5$) $\,\mu$m. {\bf (b)} Extinction ($=1-$Transmittance) spectra of unpatterned Bi$_2$Se$_3$ film (60\,nm thickness) at different temperatures. Dashed lines represent calculated extinctions for the unpatterned film (see text and Methods). {\bf (c-e)} Optical images (c), measured extinction (d), and calculated extinction (e) of Bi$_2$Se$_3$ ring arrays for different ring sizes, as specified in (c) (see labels with the values of $D$ and $d$, respectively). The array lattice period is $4d$ and the $D/d$ aspect ratio is close to 3 in all samples, whereas the ring size decreases from the top to the bottom panel. The transmittance is normalized to that of the sapphire substrate measured at the same temperature. The calculations of (e) are performed with (solid curves) and without (broken curves) the inclusion of the TI $\alpha$ phonon.}
\label{fig1-paper}
\end{center}
\end{figure*}

\section{Results}

Five films of Bi$_2$Se$_3$ (thickness $h=60$ quintuple layers $\simeq60$\,nm) were grown by molecular beam epitaxy on 0.5\,mm thick sapphire  (Al$_2$O$_3$) substrates \cite{Bansal, Bansal1}. The films were characterized through resistivity and Hall measurements \cite{Bansal, Bansal1} (see Methods for details) indicating the presence of a 2D gas of Dirac electrons with a surface density $n_{D}=3\pm1\times10^{13}$ cm$^{-2}$. One of the films was kept as grown to serve as a reference, while the other four were patterned by electron-beam lithography and reactive-ion etching to form periodic hexagonal arrays of rings with different sizes. The outer ($D$) and inner ($d$) ring diameters were  chosen to have a similar disk-to-hole aspect ratio ($D/d\simeq$ 3), while the lattice period $a$ was taken to be approximately 4 times the inner diameter ($a/d=4$) in all samples. Fig. \ref{fig1-paper}a shows a scanning electron microscope (SEM) image of the $D=15 \, \mu$m patterned film, whereas Fig. \ref{fig1-paper}c presents optical microscope images of the different arrays.

The transmittance $T(\nu)$ of the five films was measured in the terahertz domain using a Fourier-transform interferometer at temperatures held in the range of 10 to 300 K in the AILES beamline at the synchrotron Soleil \cite{pascale1, pascale2}. Each transmittance has been normalized to that of the sapphire substrate collected at the same temperature. 
The  corresponding extinction coefficient (Extinction = 1-Transmittance) is plotted in Fig. \ref{fig1-paper}b for the as-grown sample and in Fig. \ref{fig1-paper}d for the patterned films at 10\,K (blue curves), 100\,K (green curves), and 300\,K (red curves).

The extinction spectra in Fig. \ref{fig1-paper}d exhibit several peaks shifting to higher frequency (blueshifting) as the ring size is decreased. These features are accompanied by a prominent dip-peak structure at a nearly constant frequency $\sim2\ $THz, independent of ring size. Additionally, a strong dependence on temperature is observed, including a reduction in the contrast of the observed dip-peak structure, as well as a broad increase in the extinction at higher frequencies.

Quantitative insight into the origin of the observed spectra is obtained from theoretical simulations performed starting from the dielectric properties of the unpatterned film (see Methods for details). Since the thickness of the Bi$_2$Se$_3$ film is small compared with both the lateral size of the rings and the terahertz wavelengths, we can model them through a 2D surface conductivity (see Methods). Before analyzing the effects of nanostructuring, it is worth looking into the properties of this conductivity, as extracted from extinction measurements for the unpatterned film of the same thickness (Fig. \ref{fig1-paper}b). The extinction of this film exhibits a phonon peak ($\alpha$ mode) at 1.85 THz, already observed in Bi$_2$Se$_3$ single crystals \cite{DVN12}. This phonon is superimposed on an electronic background that decreases monotonically with frequency and that is associated with Dirac surface states \cite{Armitage}. The resulting conductivity is then modeled through two contributions: a Lorentzian to represent the $\alpha$ phonon and a Drude term accounting for the response of the surface Dirac electron gas (see Eq. \ref{sigma} in Methods). The former is expected to depend linearly on the film thickness, as the optical phonons are spatially localized and contribute all over the volume of the material. In contrast, the Drude term is independent of thickness but depends on the combined surface electron densities on both sides of the film. This metallic contribution is also strongly dependent on temperature (see Fig. \ref{fig1-paper}b), presumably as a result of thermal smearing of the surface electron gas (notice that $K_BT/\hbar=2.1$ THz at $T=100$ K).

By fitting the spectra of the homogeneous film (see Tab.\ref{para} in Methods), we can parametrize the TI film conductivity and use it to perform classical electromagnetic simulations as discussed 
in the Methods section. Calculations are carried out for all three temperatures under consideration and the resulting theoretical spectra very well reproduce both the experimental extinction of the unpatterned film (Fig. \ref{fig1-paper}b), and those of the patterned ones (Fig. \ref{fig1-paper}e). In particular for the ring-arrays (Fig. \ref{fig1-paper}e), analytical calculations reproduce the extinction shape, the main features of the temperature dependence noted above, as well as the general trend of the evolution of spectral maxima and minima with ring size. 

If we artificially switch off the $\alpha$ phonon contribution from the film conductivity (i.e., by removing the Lorentzian term in Eq. \ref{sigma} of Methods), the extinction spectra are simplified and consist of two dominant maxima that evolve towards higher frequencies for smaller rings (Fig. \ref{fig1-paper}e, dashed curves). Because only the Drude term contributes to the conductivity in this case, the observed features can be undoubtedly assigned to plasmonic excitations. Closer examination of the origin of these features (see Methods) reveals that they are the bonding (at low energy) and antibonding (at higher energy) ring plasmons. Moving back to the full simulation including phonons (Fig. \ref{fig1-paper}e, solid curves), we still identify the presence of both types of plasmons, although the broad bonding mode is now strongly perturbed by the relatively narrow optical phonon. The resulting plasmon-phonon interaction gives rise to characteristic dip-peak line shapes that are reminescent of the Fano interference observed in other plasmonic systems \cite{NNANO, F1961,LZM10,MFK10}.

For the smallest ring under consideration ($D=3\,\mu$m), a slight disagreement is observed at high frequency between experimental data and theory (cf. Fig. \ref{fig1-paper}d,e, bottom panels). In particular, the experimental anti-bonding absorption peak (above the $\alpha$ phonon), appears to be broader than the calculated one. This could be ascribed to unaccounted decay channels other than ohmic (single-particle) scattering, which is the only one considered in the calculation and common to both plasmons.
In this respect, fabrication defects cannot be ruled out, as they are expected to have stronger influence on the high-frequency (antibonding) plasmon mode.

\begin{figure*}
\begin{center}
\includegraphics[width=120mm,angle=0,clip]{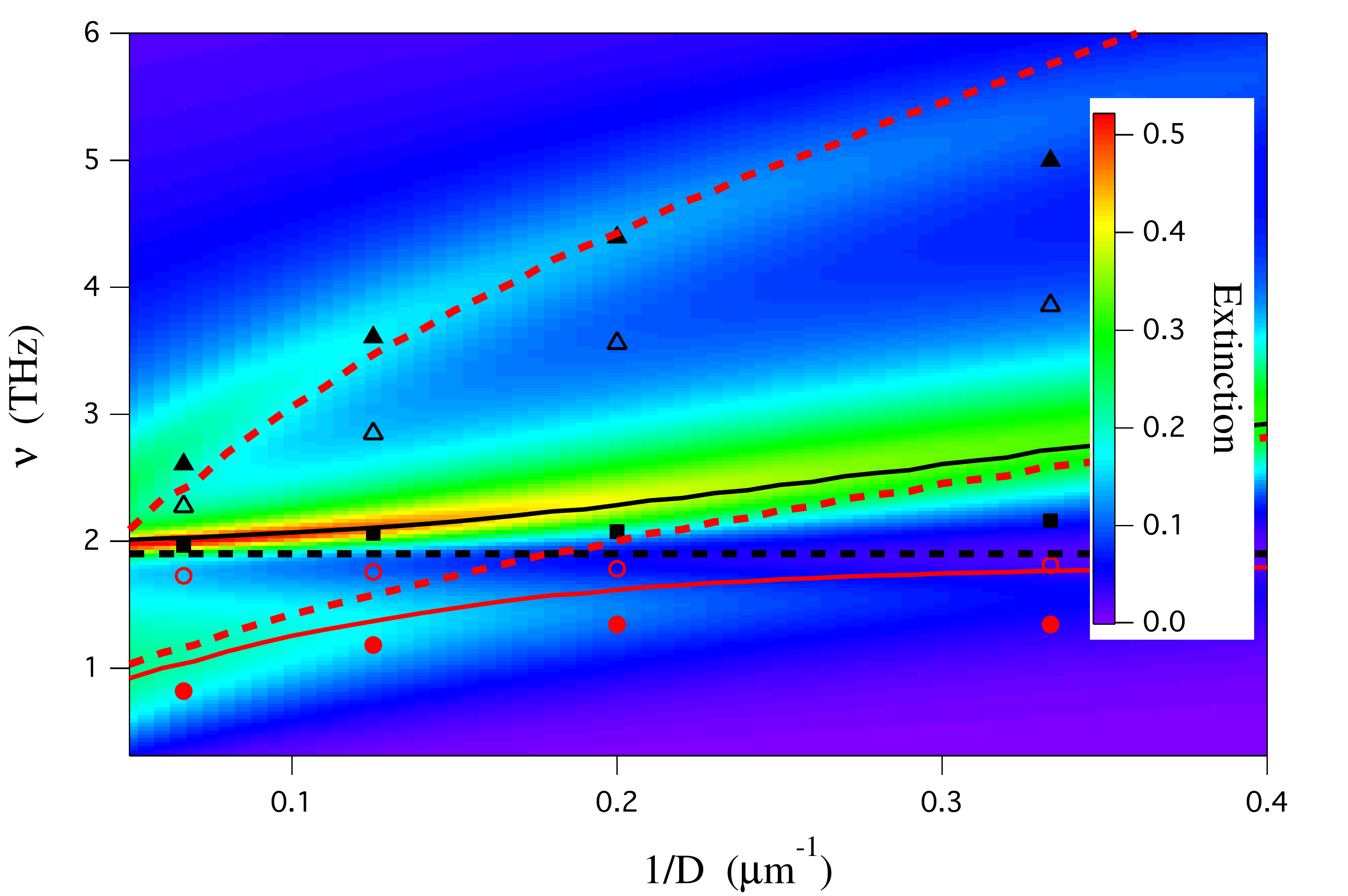}
\caption{{\bf Plasmon-phonon hybridization in Bi$_2$Se$_3$ TIRs.} We represent the extinction maxima (solid symbols) and minima (open symbols) of the TIR arrays of \ref{fig1-paper} in a dispersion diagram as a function of inverse outer diameter $1/D$ and light frequency $\nu$, superimposed on a color plot with the theoretically calculated extinction. The $\alpha$ phonon frequency of Bi$_2$Se$_3$ is shown as a horizontal black-dashed line. The two red-dashed curves represent the non-interacting plasmon frequencies calculated without including the $\alpha$ phonon (see text and Methods), while the solid curves are obtained from an effective hybridization model (see main text).}
\label{fig2-paper}
\end{center}
\end{figure*}

A broad picture of plasmon-phonon hybridization and its dependence on ring size is presented in Fig. \ref{fig2-paper}, where the background color plot shows the extinction coefficient obtained from our analytical model (see Methods), including the TI $\alpha$ phonon, as a function of inverse outer ring diameter $1/D$ and radiation frequency $\nu$. This is compared with the peak frequency of bonding (lower red-dashed curve) and antibonding (upper red-dashed curve) plasmons obtained when the $\alpha$ phonon is dismissed; the bonding mode traverses the $\alpha$ phonon frequency (black-dotted horizontal line) and these two modes hybridize, giving rise to extinction maxima above and below the phonon line (and a minimum around the bare phonon frequency). In contrast, the antibonding plasmon is far up in frequency compared to the phonon, so that it only experiences a mild redshift due to interaction with the latter. The experimentally observed maxima and minima (Fig. \ref{fig2-paper}, solid and open symbols, respectively) are in excellent agreement with the analytical calculation (color plot), including a similar hybridization of the bonding plasmon and the phonon, and a lack of hybridization of the antibonding plasmon. This bonding plasmon-phonon hybridization is a classical example of avoided crossing, which can be actually described through a simple interaction model using an effective Hamiltonian of matrix elements $\langle {\rm pl}|H|{\rm pl}\rangle=2\pi\hbar\nu_{\rm pl}$, $\langle {\rm ph}|H|{\rm ph}\rangle=2\pi\hbar\nu_{\rm ph}$, and $\langle {\rm pl}|H|{\rm ph}\rangle=\langle {\rm ph}|H|{\rm pl}\rangle=2\pi\hbar\nu_{\rm int}$, where $|{\rm pl}\rangle$ and $|{\rm ph}\rangle$ represent the plasmon and phonon modes, respectively, and $\nu_{\rm int}$ describes their interaction. Taking the plasmon and phonon frequencies $\nu_{\rm pl}$ and $\nu_{\rm ph}$ from the lower red-dashed curve and the black-dotted line of Fig. \ref{fig2-paper}, respectively, we find the following hybridized frequencies $\nu_\pm=(1/2)\left[\nu_{\rm pl}+\nu_{\rm ph}\pm\sqrt{(\nu_{\rm pl}-\nu_{\rm ph})^2+4\nu_{\rm int}^2}\right]$. These are shown as solid (red) curves in the same figure giving rise to an interaction frequency $\nu_{\rm int}=0.35\,$THz (i.e., $2\pi\hbar\nu_{\rm int}=1.4\,$meV), in excellent agreement with the full calculation (underlying color plot). It should be noted that this avoided crossing involves a relatively narrow mode (the phonon), which can be regarded as a discrete excitation, and a broad one (the bonding plasmon), which acts as a continuum, thus leading to Fano interference \cite{F1961,LZM10,MFK10}, clearly supported by the presence of a deep dip placed in between the two frequencies of the resulting hybridized states (see Fig. \ref{fig2-paper}).

In summary, we here report the observation of plasmons in arrays of rings carved in a topological insulator (Bi$_2$Se$_3$) film, as well as their hybridization with a prominent optical phonon of this material in the THz regime. As the structures have a thickness that is small compared with their lateral dimensions, we are able to describe them through classical electrodynamics, using the experimental frequency-dependent surface conductivity of the uniform film as input. The latter is taken as the sum of a phonon Lorentzian term describing the $\alpha$ Bi$_2$Se$_3$ phonon, and a Drude term related to Dirac topological surface states, with parameters as obtained from a fit to optical measurements of unpatterned films. The good agreement that we find between theory and experiment through this procedure suggests that the response is mainly local. Furthermore, the rings are small compared to the light wavelength, which allows us to describe them as dipolar scatterers in order to account for the combined effect in the arrays. A characteristic bonding/antibonding plasmon-mode pattern is observed in the ring response, which is strongly affected by the $\alpha$ TI optical phonon, leading to Fano profiles in the measured extinction spectra. We further observe a strong temperature dependence, we attribute to thermal smearing of the electron distribution in the Dirac-fermion states at the insulator surfaces. This behavior strongly supports exploiting topological insulator plasmonic resonances in the design of thermally tunable terahertz devices.

\section{Methods}

{\bf Sample fabrication.} High quality Bi$_2$Se$_3$ thin-films of 60 quintiple layer (QL) thickness (1 QL$\simeq$1 nm) were prepared by a custom-designed SVTA MOS-V-2 Molecular Beam Epitaxy (MBE) system using the standard two-step growth method \cite{Bansal, Bansal1}. The $10\times10\,$mm$^{2}$ Al$_2$O$_3$ substrates were first cleaned by heating up to 750\,$^\circ$C in an oxygen environment to remove organic surface contamination. The substrates were then cooled and kept at 110\,$^\circ$C for an initial deposition of 3 quintuple layers of Bi$_2$Se$_3$. This was followed by heating to 220\,$^\circ$C, at which the remainder of the film was deposited to reach the targeted thickness. The Se and Bi flux ratio was kept constant to approximately Se:Bi$=$10:1 in order to minimize Se vacancies. Once the films were cooled, they were removed from the vacuum chamber and vacuum-sealed for transport and handling.

Bi$_2$Se$_3$ rings were patterned by electron-beam lithography (EBL) and subsequent Reactive Ion Etching (RIE). The Bi$_2$Se$_3$ film was spin-coated with a double layer of electron-sensitive resist polymer PMMA (Poly-(methyl methacrylate)) up to a total thickness of 1.4 microns. The hexagonal ring  patterns with different ring sizes were then written in the resist by EBL. In order to obtain a lithographic pattern with re-alignment precision below 10 nm over a sample area suitable for THz spectroscopy ($\sim10\times10\,$mm$^2$), we used an electron beam writer equipped with a XY interferometric stage (Vistec EBPG 5000). The patterned resist served as a mask for the removal of Bi$_2$Se$_3$ by RIE at low microwave power of 45\,W to prevent heating of the resist mask.  Sulfur hexafluoride (SF$_6$) was used as the active reagent. The Bi$_2$Se$_3$ film was etched at a rate of 20\,nm/min, and subsequently verified by atomic force microscopy (AFM) after soaking the sample in acetone to remove PMMA. The in-plane edge quality after the RIE process, as inspected by AFM, closely followed that of the resist polymer mask (i.e., edge roughness smaller than 20\,nm). 

{\bf Optical characterization.} The THz absorption spectra were acquired by using a Bruker IFS-125 Michelson interferometer and a liquid-He cooled Infrared-Lab bolometer combined with synchrotron radiation at the AILES beamline at Synchrotron SOLEIL.  The Bi$_2$Se$_3$ film and a co-planar Al$_2$O$_3$ bare substrate were mounted on the cold finger of a pulsed-tube cryoogenerator (Cryomec) and kept at a pressure of about 10$^{-6}$\,mbar. The extinction coefficient  reported in \ref{fig2-paper} corresponds to $1-T$, where $T$ is the film transmittance, defined as the ratio between the intensity transmitted through the thin film and that transmitted through the bare substrate.

{\bf Modeling the optical response of a topological insulator thin film.} We intend to simulate the optical response of a periodically patterned topological insulator film of thickness $h$ small compared with both the lattice period $a$ and the lateral dimensions of each structure. Under these conditions, the response of the film can be described through its effective surface conductivity $\sigma(\omega)$, which we assume to be a function of radiation frequency $\omega$ only. We thus neglect nonlocal effects, which should only play a minor role for objects of large lateral dimensions compared with the Fermi wavelength of the electron gas in the film. We are describing a spectral region dominated by the presence of an optical phonon, so that its bulk dielectric function can be approximated as \cite{Dressel}
\begin{equation}
\epsilon_{\rm ti}(\omega)=\epsilon_\infty+\frac{S^2_{\rm ph}}{\omega_{\rm ph}^2-\omega(\omega+\ii\gamma_{\rm ph})},
\label{epsti}
\end{equation}
where $\omega_{\rm ph}$ is the phonon frequency, $\gamma_{\rm ph}$ is the phonon width, and $\epsilon_\infty$ is the high-$\omega$ dielectric function. \ref{epsti} can be equivalently expressed in terms of a bulk conductivity $\sigma^{\rm 3D}$ as $\epsilon_{\rm ti}=1+4\pi\ii\sigma^{\rm 3D}/\omega$, and consequently, it contributes to the 2D surface conductivity of the film $\sigma$ with a term $h\sigma^{\rm 3D}$, proportional to the film thickness $h$. The two conducting regions on either side of the film produce an additional contribution that should take the form of a Drude term at low frequencies. In summary, we can write the 2D conductivity of the topological insulator thin film as
\begin{equation}
\sigma(\omega)=\frac{e^2}{\hbar}\frac{\ii\omega_{\rm e}}{\omega+\ii\gamma_{\rm e}}+\frac{\ii\omega h}{4\pi}\frac{S^2_{\rm ph}}{\omega(\omega+\ii\gamma_{\rm ph})-\omega_{\rm ph}^2}+\frac{\ii\omega h}{4\pi}(1-\epsilon_\infty),
\label{sigma}
\end{equation}
where $\gamma_{\rm e}$ is the intrinsic decay rate of its surface electronic excitations, whereas $\omega_{\rm e}$ is a characteristic frequency that should scale with the $1/4$ power of the electron gas real density, given the Dirac-fermion character of the latter. We note that $\omega_{\rm e}$ should rather be independent of dielectric environment and film thickness $h$, provided there is no overlap between the two metallic regions of the topological insulator. 
%Incidentally, the Drude term of \ref{sigma} also describes intraband screening in graphene, with %$\omega_{\rm e}=E_{\rm F}/\hbar$ expressed in terms of the Fermi energy $E_{\rm F}$.

We extract the parameters of the above model for the conductivity by means of a fit of the extinction spectra of the unpatterned samples, using the standard formula for thin films on a substrate \cite{tinkham}
\begin{equation}
E(\omega)=1-\frac{1}{[1+Z_0\sigma_1(\omega)/(n+1)]^2+[Z_0\sigma_2(\omega)/(n+1)]^2},
\label{tinkham}
\end{equation}
where $Z_0=377\, \Omega$ is the free-space impedance, $\sigma_{1}(\omega)$ ($\sigma_{2}(\omega)$) is the real (imaginary) part of the conductivity, and $n$ is the refractive index of the substrate ($n=3.2$ for Al$_2$O$_3$ in the THz range). The resulting parameters are collected in \ref{para}. The effect of these parameters is described by the self-consistent optical response elaborated in next paragraph.

\begin{table*}[htbp]
\centering
\begin{tabular}{r|ccccc}
          &  $\omega_{\rm e}/2\pi$ (THz)  & $\gamma_{\rm e}/2\pi$ (THz) & $S_{\rm ph}/2\pi$ (THz)  & $\omega_{\rm ph}/2\pi$ (THz) & $\gamma_{\rm ph}/2\pi$ (THz) \\
          \hline
10\,K   &  34.1 & 1.4 & 20.6 & 1.9 & 0.1 \\
100\,K &  42.3 & 2.7 & 21.2 & 1.9 & 0.1 \\
300\,K &  51.0 & 4.3 & 20.6 & 2.0 & 0.2 \\
\end{tabular}
\caption{\bf Fitting parameters extracted to model the conductivity of unpatterned films according to \ref{sigma}.}
\label{para}
\end{table*}

{\bf Calculation of the transmittance of a periodic thin-ring array.} We consider thin scatterers of small lateral size $D$ arranged in a periodic lattice of period $a$, placed at the interface between materials of refractive indices $n_1=\sqrt{\epsilon_1}$ and $n_2=\sqrt{\epsilon_2}$. For simplicity, we focus on normally incident light coming from medium 1. We follow a previously reported formalism to obtain the  transmission coefficient across this decorated interface as \cite{paper182,paper212}
\begin{equation}
t=t_0\left(1+\frac{\ii S}{\alpha^{-1}-G}\right),
\label{t}
\end{equation}
where $t_0=2n_1/(n_1+n_2)$ is the transmission coefficient for the bare interface, $G=2g/[a^3(\epsilon_1+\epsilon_2)]+\ii S$ accounts for inter-site interactions in the array, $g=5.52$ for a hexagonal lattice,
$S=(4\pi\omega/cA)/(n_1+n_2)$, $A$ is the unit-cell area (i.e., $A=(\sqrt{3}/2)a^2$ for a hexagonal lattice), $\omega$ is the light frequency, and $\alpha$ is the polarizability of each scatterer including the effect of the interface (i.e., the actual induced dipole is given by $\alpha$ times the sum of the external field plus the fields scattered by the planar interface and the rest of the scatterers). Following an electrostatic scaling law that is applicable in the $D\omega/c\ll1$ limit, we find \cite{paper235}
\begin{equation}
\alpha=D^3\sum_j\frac{A_j}{(2L_j)/(\epsilon_1+\epsilon_2)-\ii\omega D/\sigma},
\label{alpha}
\end{equation}
where $A_j$ and $L_j$ are material- and size-independent constants that run over resonant modes $j$, whereas $\sigma$ is the 2D conductivity of the thin film (see \ref{sigma} above). This formalism is valid under the condition $a\omega/c\ll1$, which is fulfilled in this work.

The extinction is calculated here as $1-|t/t_0|^2$ where $t_0$ is the transmission coefficient of the sapphire substrate. For simplicity, \ref{t} is given for normal incidence, although the calculations here reported are mildly corrected by averaging over a finite range of incidence angles ($|\theta|<\sin^{-1}{\rm NA}$ for the finite numerical aperture ${\rm NA}=0.11$ used in our experiments), using an extension of the present theory to oblique incidence \cite{paper235}.

{\bf Parameters for Bi$_2$Se$_3$ rings.} We consider rings of inner (outer) diameter $d$ ($D$). These structures are characterized by a bonding mode as well as an antibonding mode at higher frequency. In particular, we present calculated results in Fig. \ref{fig1-paper},\ref{fig2-paper} for hexagonal arrays of rings with a $D=3d$ aspect ratio, characterized by the polarizability of \ref{alpha} with only two terms: \cite{paper212} $A_1=0.42$ and $L_1=10.3$ for the bonding mode; and $A_2=0.17$ and $L_2=47$ for the antibonding one. The film thickness is $h=60\,$nm, the lattice period is $a=4d$, and the surrounding refractive indices are $n_1=1$ (air) and $n_2=3.2$ (Al$_2$O$_3$). The conductivity of the rings material (Bi$_2$Se$_3$) is modeled through \ref{sigma} with temperature-dependent parameters as shown in Tab. \ref{para} and taking $\epsilon_\infty=30$.

\balancecolsandclearpage

\section{Author Information}
\subsection{Corresponding Author}
*E-mail: stefano.lupi@roma1.infn.it (S.L.); javier.garciadeabajo@nanophotonics.es (J.G.A).

\subsection{Author Contributions} 
M. B., N. K. and S. O. fabricated and characterized Bi$_{2}$Se$_{3}$ films. A.D.G. and V.G. performed EBL lithography and etching. M.A., F.D., O.L., and S.L. carried out the terahertz experiments and data analysis. J.G.A. performed the theoretical calculations. S. L. planned and managed the project with inputs from all the co-authors.  J.G.A. and S.L. wrote the manuscript. All authors extensively discussed the results.

\subsection{Notes} 
The authors declare no competing financial interest.

\section{Acknowledgments} This work has been supported in part by the European Commission (Graphene Flagship CNECT-ICT-604391 and FP7-ICT-2013-613024-GRASP).

\noindent M.B., N.K., and S.O. are supported by National Science Foundation (NSF DMR-0845464) and Office of Naval Research (ONR N000141210456).

\noindent We acknowledge SOLEIL for providing terahertz synchrotron radiation.

\section{References}
\bibliography{NC_refs}

\end{document}